\documentclass[11pt,a4paper]{article}
\usepackage{jcappub}

 \usepackage{amsmath}
 \usepackage{float}
 \usepackage{placeins}
\usepackage{epsfig}

 \usepackage{graphicx}
 \usepackage{amssymb}
 \usepackage{amsfonts}
 \usepackage{latexsym}
 \usepackage[latin1]{inputenc} 

\renewcommand{\vec}[1]{{\bf #1}}

\def\beq{\begin{eqnarray}}
\def\eeq{\end{eqnarray}}
\def\bea{\begin{eqnarray}}       
\def\eea{\end{eqnarray}}

\def\al{\alpha}
\def\be{\beta}

\def\ga{\gamma}
\def\de{\delta}

\def\ep{\epsilon}

\def\ka{\kappa}
\def\la{\lambda}
\def\na{\nabla}
\def\pa{\partial}

\def\si{\sigma}
\def\om{\omega}

\def\th{\theta}

\def\Ga{\Gamma}

\title{Dirac fields, torsion and Barbero-Immirzi parameter
in Cosmology}

\author[a]{G. de Berredo-Peixoto}
\author[b]{L. Freidel}
\author[c]{I. L. Shapiro}\note{On leave from Tomsk State
Pedagogical University, Tomsk, Russia.}
\author[d]{C. A. de Souza}

\affiliation[a]{Departamento de F\'{\i}sica, ICE,
Universidade Federal de Juiz de Fora, MG, Brazil}
\affiliation[b]{Perimeter Institute for Theoretical Physics
31 Caroline St. N, N2L 2Y5, Waterloo ON, Canada}
\affiliation[c]{Departamento de F\'{\i}sica, ICE,
Universidade Federal de Juiz de Fora, MG, Brazil}
\affiliation[d]{Departamento de F\'{\i}sica, ICE,
Universidade Federal de Juiz de Fora, MG, Brazil}

\emailAdd{guilherme@fisica.ufjf.br}
\emailAdd{lfreidel@perimeterinstitute.ca}
\emailAdd{shapiro@fisica.ufjf.br}
\emailAdd{abrahaocleber@gmail.com}

\abstract{
We consider cosmological solution for Einstein gravity with massive
fermions with a four-fermion coupling, which emerges from the Holst
action and is related to the Barbero-Immirzi (BI) parameter. This
gravitational action is an important object of investigation
in a non-perturbative formalism of quantum gravity. We study the
equation of motion for the Dirac field within the standard
Friedman-Robertson-Walker (FRW) metric. Finally, we show  the theory
with BI parameter and minimally coupling Dirac field, in the zero mass
limit, is equivalent to  an additional term which looks like a perfect
fluid with the equation of state \ $p = w\rho$, with $w = 1$ which is
independent of the BI parameter. The existence of mass imposes a
variable $w$, which creates either an inflationary phase with $w=-1$,
or assumes  an ultra hard equation of states $w = 1$ for very early
universe. Both phases relax to a pressure less fluid  $w = 0$ for
late universe (corresponding to the limit $m\to\infty$).}

\keywords{Dirac fields, Cosmology, Barbero-Immirzi parameter}

\begin{document}
\maketitle


\section{Introduction}

The investigation of Dirac fields in curved spacetime has been
addressed in several works for many years, especially
in the last decade (see, for example, Refs.
\cite{deser,epstein,kremer2,cheng,boeckel,
chimentokremer,samojeden,rakhi,rakhi2,fabbri}).
There are many papers considering not only the classical aspects
of the theory, but also quantum ones. In the present work, we deal
with the classical aspects of massless Dirac particles
in the spatially flat Friedmann-Robertson-Walker background in the
presence of the Barbero-Immirzi parameter \cite{barbero,immirzi},
refers to as BI parameter in the following. This parameter,
described by the Holst action term \cite{holst}, was introduced
under the non-perturbative quantum gravity perspective, and it
represents then a new dimensionless parameter coming from
a more fundamental theory.

The vacuum gravitational action together with Holst term can be
written as
\beq
S_H = \frac{1}{\ka}\int d^4x \sqrt{-g}\left\{
\,- \,R \,+ \, \frac{1}{\be} \,
\ep_{\al\be\mu\nu} \,R^{\al\be\mu\nu} \right\}\,,
\eeq
where $\ka = 16\pi G$, \ $\ep_{\al\be\mu\nu}$ \ is the Levi-Civita
tensor and $\be$ is the BI parameter, which should be real and
positive. In General Relativity (GR) this parameter makes no effect
on dynamical equations, because $\ep_{\al\be\mu\nu} R^{\al\be\mu\nu}$
vanishes due to the cyclic symmetry of the Riemann tensor,
$R^{\al\be\mu\nu} + R^{\al\nu\be\mu} + R^{\al\mu\nu\be} = 0$.
However, this is not so if torsion is present. The simplest way
to introduce torsion is to consider the Einstein-Cartan action coupled
to a Dirac field, such that the equations of motion for torsion become
non-trivial. For an introduction and review on the theories with torsion,
one can see \cite{revhehl,torsi,Puetzfeld}, the last reference gives
nice overview of the relevant cosmological models. It is worth noticing
that, in this scenario, where fermionic matter is present the BI
parameter and/or torsion can affect the gravitational dynamics,
providing an interesting way to investigate its classical and quantum
effects. It was first proposed in Ref. \cite{perezrovelli} that the
Immirzi parameter leads to an effective four-fermion interaction
mediated by the Immirzi parameter. The same effect can be achieved in
the framework of Einstein-Cartan theory with minimal or nonminimal
coupling of fermions to torsion.

The first studies of the effect of the Immirzi parameter was done
using the minimal coupling procedure (MCP) of the fermions to gravity
\cite{perezrovelli, freidel}. However it was pointed out first in
\cite{freidel} that the effective four-fermions interaction depends
on the choice of the coupling of fermions to gravity and that we can
assume non minimal coupling procedure.

This is due to the fact that the action describing Dirac theory in flat
spacetime is invariant by the introduction of a total divergence,
say, $\partial_\mu V^\mu$ with $V^\mu= \bar{\psi}\gamma^\mu\psi$ or
$V^\mu= \bar{\psi}\gamma^\mu\gamma_{5} \psi$. However when torsion
is present, there may be an extra coupling, for such terms produces
extra term $T_\mu V^{\mu}$ in the action
(here $T_\mu= T^{\alpha}_{\,\,\,\mu\alpha}$ is the trace of the
torsion tensor). It means that a whole class of equivalent actions
in flat spacetime are related (though MCP) to inequivalent actions
in curved spacetime. This problem was especially addressed by
Kibble \cite{kibble} (see also \cite{torsi}).

In \cite{freidel} a one complex parameter family of non minimal coupling
was investigated, in \cite{mercuri} another family of non-minimal coupling
was proposed while in \cite{alexandrov} a general  3  parameter family of
non minimal coupling was investigated. In our note we {\it assume } that
the fermion coupling is the minimal coupling and we investigate the
consequences of this coupling on the cosmological evolution.

It can be argued that the minimal coupling is the most natural coupling
of gravity to fermion based on the fact that non minimal couplings
are sourced by components of the torsion that do not appear naturally
in models of spinning matter. For instance it is possible to treat
fermions, at a large scale, as a fluid with an intrinsic spin
density. Following the Lagrangian formalism for this exotic
spin fluid \cite{weyssenhoff,ray,obukhov}, one can see that
the contribution from such fluid always produces a traceless
torsion \cite{ourpaper} (for the extensive discussion of the
theories with spin fluid in cosmology, see also  Refs.
\cite{gasperini,krawiec,Puetzfeld2,lasenby,ourpaper2,becattini,poplawski}).
In this case the trace of the
torsion tensor $T_\mu = T^\al_{\,\,\,\mu\al}$  has smaller
physical meaning and its presence is completely due to the
nonminimal procedure \cite{revhehl,torsi}. It turns
out that the non-minimal procedure \cite{bush-85}
\beq
\pa_\mu\to \na_\mu + i\eta_1\ga^5S_\mu+i\eta_2T_\mu
\label{eta}
\eeq
\ (we use the notations of \cite{torsi}, correspondingly $\na_\mu$
is the Riemannian covariant derivative, without torsion)
is the most consistent one, especially at the quantum level.
Of course it is possible to view this
procedure as a MCP with a modified connection \cite{kasmbrief}.
In any case, the result will be the equivalent action with
current-current interaction found in \cite{perezrovelli}.
For the corresponding discussion of the issue related to the BI
parameter effects, see, for example, Refs.
\cite{kasmbrief,kasm1,kasm2}.

In the present paper we consider a particular cosmological
solution, with a FRW-like metric, within the standard approach
to BI parameter, and show that this parameter does affect the
cosmological solution.

The paper is organized as follows. In the next section we
briefly discuss the existing ambiguity of the MCP in
Einstein-Cartan theory. In Sect. 3 the details of the
derivation of equations for the metric and for the
components of the Dirac field are given. The details of
the derivation of the Energy-Momentum Tensor are settled
in the Appendix A. In Sect. 4 we
discuss the cosmological solution and verify the consistency
conditions. The calculations are not completely trivial and
hence we present them in some details. Finally, in Sect. 5
we draw our conclusions and discuss possible perspectives
of cosmology based on the gravity theory with BI parameter.

\section{Derivation of dynamical equations}

The action for gravity theory with an additional Holst term
with the BI parameter and minimal fermion coupling, in the
presence of massive Dirac fields can be written in terms of a
spin-spin coupling in the form \cite{perezrovelli, freidel}
\beq
S &=& \int d^4x \sqrt{-g}\left\{
- \frac{1}{\ka}\,R + \frac{i}{2}
\big(\bar{\psi}\ga^\mu\na_\mu \psi - \na_\mu\bar{\psi}\ga^\mu\psi\big)
- m\bar{\psi}\psi - \theta J_\mu J^\mu \right\}\, ,
\label{action}
\eeq
where $\theta$ is the spin-spin
parameter related to Immirzi parameter $\be$ by
\beq
\theta = \frac{3\ka}{32}\frac{\be^2}{\be^2 + 1}
\label{th}
\eeq
and the axial current is defined as
\beq
J^\mu = \bar{\psi}\Gamma_5\ga^\mu\psi\,.
\label{J}
\eeq
Here, $\Gamma_5 = i\Gamma_0\Gamma_1\Gamma_2\Gamma_3$ is the chiral
Dirac matrix. The Dirac matrices, $\ga^\mu$, are expressed in terms
of the vierbeins $e^a\mbox{}_\nu$ and the Dirac matrices $\Ga^a$ in
tangent flat space according to $\ga^\mu = e_a\mbox{}^\mu\Ga^a$.

Two important observations are in order. The same action
(\ref{action}) can be seen as the result of integrating out
completely antisymmetric torsion in the Einstein-Cartan gravity
coupled to the fermion fields forming axial current. In this
sense all our further consideration can be attributed to one of
the two models, namely Holst gravity with BI parameter and
Einstein-Cartan gravity. Further aspects of the difference
between these two formulations were discussed in \cite{mercuri}.

The second observation concerns the relevance of the choice of the
MCP procedure, which we have already mentioned in the Introduction.
The MCP corresponds to the value $\eta_1=1/8$ in (\ref{eta}). Taking
into account the relation of our model with the Einstein-Cartan
gravity, it becomes obvious that the transition to an arbitrary
non-minimal interaction between fermion and torsion is performed
by the multiplication of $\theta$ in (\ref{th}) by a factor of
\ $64\eta_1^2$. Therefore the ``nonminimal'' value of the parameter
should be
\beq
\theta_{\rm nonminimal}
\,=\, \frac{6\ka\,\eta_1^2\,\be^2}{\be^2 + 1}\,.
\label{th-nonmi}
\eeq
In the ideal world with only one kind of non-interacting
fermions, this would be the end of the story, because the single
parameter $\theta$ contains the information about both $\beta$
and $\eta_1$. However, in many physically interesting theories
such as Standard Model of particle physics, there are different
kinds of fermions,
all of them  interacting to scalar (Higgs) by means of Yukawa
interaction. At quantum level the nonminimal parameters $\eta_1$
for different fermions are renormalized, become running parameters,
and their running depends on the values of Yukawa couplings
\cite{bush-85}. As a consequence, the values of $\eta_1$ must be
different for different fermions. In what follows we present a
simplified consideration with only one $\theta$, corresponding
to the one single type of fermionic axial current $J^\mu$. One can
regard this approach as simplification, which is based on the
assumption that the running of different parameters $\eta_1$ is
not very strong and that MCP hypothesis can serve, after all,
as a good approximation.

The variation with respect to the Dirac fields gives the dynamical
equations
\beq
i\ga^\mu\na_\mu\psi
&=&  2\theta J^\mu\Gamma_5\ga_\mu\psi + m\psi \,,
\label{psi1} \\
i\na_\mu\bar{\psi}\ga^\mu
&=&  -\,2\theta J^\mu\bar{\psi}\Gamma_5\ga_\mu - m\bar{\psi} \,.
\label{psi2}
\eeq

The Einstein equations, coming from the variation with respect
to \ $g_{\al\be}$, \ can be written as\footnote{The expression
for the Energy-Momentum Tensor for the free case is known from
\cite{BW-1957}.}
\beq
R_{\al\be} - \frac12\,R\,g_{\al\be}
\,=\, G_{\al\be} \,=\, \frac{\ka}{2} T_{\al\be}
= \frac{i}{4}\,\ka\, \big[
\bar{\psi}\ga_{(\al}\na_{\be )}\psi
- \na_{(\be}\bar{\psi}\ga_{\al )}\psi \big] - \frac{1}{2}\ka g_{\al\be}
{\cal L}\,,
\label{E}
\eeq
where
\beq
{\cal L} = \frac{i}{2}(\bar{\psi}\ga^\mu\na_\mu \psi -
\na_\mu\bar{\psi}\ga^\mu \psi) - m\bar{\psi}\psi - \theta J_\mu J^\mu\,,
\label{L}
\eeq
and \ $\ga_{(\al}\na_{\be )}$ \ means
$\frac{1}{2}(\ga_{\al}\na_{\be} + \ga_{\be}\na_{\al})$.
By using equation (\ref{psi1}), we obtain
\beq
G_{\al\be} = \frac{i}{4}\,\ka\,\left[\bar{\psi}\ga_{(\al}\na_{\be )}\psi
- \na_{( \be}\bar{\psi}\ga_{\al )}\psi\right]
- \frac{\theta}{2}\ka g_{\al\be} J_\mu J^\mu\,.
\label{einstein}
\eeq

Let us note that our expression for the energy-momentum tensor
(\ref{E}) is slightly different from the one of \cite{deser},
which was obtained by variation with respect to the vierbein.
The expression (\ref{E}) has been obtained by variating with
respect to the metric. The technical details can be found in
the Appendix A.

The next step is to calculate the equations for the specific
metric corresponding to the homogeneous and isotropic universe
(FRW space-time). The spin connection,
\beq
\omega^{ab}\mbox{}_\mu & = & \frac{1}{4}\left(
e^{b\al}\partial_\mu e^a\mbox{}_\al
- e^{a\al}\partial_\mu e^b\mbox{}_\al +
e^{a\al}\partial_\al e^b\mbox{}_\mu
- e^{b\al}\partial_\al e^a\mbox{}_\mu +
\right. \nonumber \\
& + & \left. e^{b\nu} e^{a\la} e_{c\mu}\partial_\la e^c\mbox{}_\nu
-e^{a\nu} e^{b\la} e_{c\mu}\partial_\la e^c\mbox{}_\nu \right)\,,
\eeq
can be obtained for the FRW spacetime with null spatial curvature,
$ds^2 = dt^2 - a^2(t)(dx^2 + dy^2 + dz^2) =
a^2(\eta)(d\eta^2 - dx^2 - dy^2 - dz^2)$, by taking into account
the vierbeins
\beq
e^a\mbox{}_0 = (a(\eta),0,0,0)\,,\;\;\;
e^a\mbox{}_1 = (0,a(\eta),0,0)\,,\;\;\;
e^a\mbox{}_2 = (0,0,a(\eta),0)\,,\;\;\;
e^a\mbox{}_3 = (0,0,0,a(\eta))\,.
\label{tetrad}
\eeq
Here we consider the conformal time, $\eta$, defined by $a(\eta)d\eta = dt$.
Thus, the non-null spin connection components $\omega^{ab}\mbox{}_\mu$ are
\beq
\omega^{01}\mbox{}_1 = \omega^{02}\mbox{}_2 = \omega^{03}\mbox{}_3 =
\frac{a^\prime}{2a}\,, \;\;\;\;\;\;\;
\omega^{10}\mbox{}_1 = \omega^{20}\mbox{}_2 = \omega^{30}\mbox{}_3 =
-\frac{a^\prime}{2a}\,,
\label{connect}
\eeq
where $a^\prime = da/d\eta$.

\section{Cosmological solution and consistency conditions}
\subsection{Dirac equation}

In order to solve the equation in a cosmological background, we
make the assumption that the fermions are spatially constant,
$\partial_{i}\psi=0$. With this assumption and using the form
of the connection (\ref{connect}) we get
\beq \label{der}
\na_0\psi = \pa_0\psi
\,,\quad
\na_0\bar{\psi} = \pa_0\bar{\psi}
\,,\quad
\na_i\psi = \frac{a^\prime}{2a}\Ga_i \Ga_0\psi
\,,\quad
\na_i\bar{\psi} = \frac{a^\prime}{2a}\bar{\psi}\Ga_0\Ga_i
\,,\quad
i = 1,2,3.
\eeq

The Dirac equations (\ref{psi1}) and (\ref{psi2}) can be written as
\beq
\frac{i}{a} \,\Ga_0\Big(\psi^\prime + \frac{3a^\prime}{2a}\psi \Big)
& = &
 m\psi
+ 2 \theta \, (\Ga_5\Ga_{a} \psi ) J^a
\nonumber
\\
-\,\frac{i}{a} \,\Big(
\bar{\psi}^\prime + \frac{3a^\prime}{2a}\bar{\psi} \Big) \Ga_0
& = &
m\bar{\psi} + 2 \theta\, J^a(\bar{\psi}\Ga_5\Ga_{a})  \,.
\eeq
where $J_a \equiv (\bar{\psi}\Ga_5\Ga_a\psi )$.

From these equation we can extract the conservation law:
\beq
(\bar{\psi}\Ga_0\psi)^\prime = -\frac{3a^\prime}{a} (\bar{\psi}\Ga_0\psi)\,.
\label{dirac eq}
\eeq
which is identical to the
current conservation
\beq
0 = \na_\mu (\bar{\psi}\ga^\mu\psi) = \pa_\mu (\bar{\psi}\ga^\mu\psi)
+ \Ga^\mu\mbox{}_{\rho\mu} (\bar{\psi}\ga^\rho\psi)\,,
\eeq
where $\Ga^\mu\mbox{}_{\rho\mu}$ is the Riemannian affine connection.
This can be easily integrated out
$$
{V}_{0}(\eta) = \frac{M^{3}}{a^{3}}
$$
where $M$ is a mass scale that gives the size of the condensate and
$V_{a}\equiv \bar\psi \Gamma_{a} \psi$ is the vectorial current.

One can also get the following equations for the symplectic current:
\beq
\frac{i}{2a}\left( \bar\psi \Gamma_{0}\psi^{\prime}- \bar\psi^{\prime}
\Gamma_{0}\psi\right) &=& 2\theta J_{a} J^{a} + m \bar\psi \psi,
\label{sym1}
\\
\frac{i}{2a}\left( \bar\psi \Gamma_{i}\psi^{\prime}- \bar\psi^{\prime}
\Gamma_{i}\psi\right) &=& 0.
\label{sym2}
\eeq

In order to analyze further the spinor equations of motion, let us
introduce  not only the vectorial and axial current $V_a$ and $J_a$
but also the pseudo scalar
$P\equiv i \bar\psi \Ga_{5}\psi$, scalar  $S\equiv \bar\psi \psi$ and
Lorentz  tensor $L_{ab}= \frac{ i}{2} \bar\psi [\Ga_{a},\Ga_{b}]\psi$
currents.

We can check that the equation of motion for the scalar $S$ and
pseudo scalar $P$ and the time component of the axial current form
a closed subset of equation. Indeed, in the Appendix B we derived
the closed set of equations
\beq
\label{direq1}
\frac{1}{2a}\,(a^{3} S)^{\prime} &=& 2 a^{3} \theta\, P J_{0}\,,
\\
\label{direq2}
\frac{1}{2a}\,(a^{3} J_{0})^{\prime} &=& - a^{3} m\, P\,,
\\
\label{direq3}
\frac{1}{2a}\,
(a^{3} P)^{\prime} &=& a^{3}( m\, J_{0} -2\theta S J_{0})\,.
\eeq
From these equation we can conclude that
$a^{6}(S^{2}+J_{0}^{2}+ P^{2})$ is a conserved quantity.
Moreover we have the continuity equation:
\bea
\left[a^{3} ( \theta J_{0}^{2} + m S) \right]^{\prime}
+ 3 a^{\prime}a^{2}\theta J_{0}^{2} = 0\,.
\eea
We can also show that the spatial components of the axial vector are
conserved,
\bea
(a^{3} J_{i})^{\prime} = 0\,.
\eea

For completeness we can also give the evolution equation for the
other currents:
\beq
\frac{1}{2a^4}\,\left(a^{3} V_{i} \right)^{\prime}
&=&  mL_{0i}+ 2 \theta
\epsilon_{iaj} J^{a} V^{j}\,,
\nonumber
\\
\frac{1}{2a^4}\, (a^3 L_{0i})^{\prime}
&=& - m V_{i} + 2\theta \epsilon_{iab}
J^{a}L^{0b} - \theta J_{0}\epsilon_{ijk} L^{jk}\,,
\nonumber
\\
\frac{1}{2a^4}\,(a^{3} L^{i})^{\prime}
&=& 2\theta\big(\epsilon^{jki}
J_{j}L_{k}  -  2J_{0}  L^{0i}\big)\,,
\label{mais curts}
\eeq
where $L^{i}\equiv 1/2 \epsilon^{ijk}L_{jk}$, and
$\epsilon_{aij}$ is the 3-dimensional Levi-Civita tensor
($\epsilon_{aij} = 0$ if $a=0$).
One can extract from these equations one conserved quantity, namely
\beq
a^{6}( V_{i} V^{i}
+ L^{0i}L_{0i} + L_{i}L^{i})\,.
\eeq

\subsection{Einstein equation}
The spatial and time-space  off-diagonal components of the fermion
energy momentum tensor are given by
\beq
T_{ij} & = & \frac{i}{2}\left( \bar{\psi}\ga_{(i}\na_{j )}\psi
- \na_{(j}\bar{\psi}\ga_{i)}\psi\right)
+ a^{2} \eta_{ij} \theta J^{\mu}J_{\mu}= 0\,,\;\;\;\; i\neq j\,,
\label{cons1}
\\
T_{0i} & = & \frac{i}{2}\left( \bar{\psi}\ga_{(0}\na_{i )}\psi
- \na_{(0}\bar{\psi}\ga_{i)}\psi\right)\,,
\label{cons2}
\\
T_{00} &= &  \frac{i}{2}\left( \bar{\psi}\ga_{0}\na_{0}\psi
- \na_{0}\bar{\psi}\ga_{0}\psi\right) - a^{2} \theta J^{\mu}J_{\mu}\,.
\label{cons3}
\eeq
Direct algebraic manipulation with Eq. (\ref{cons1}), using the
derivative (\ref{der}) together with the identity
$\Ga_{(i}\Ga_{j)}\Ga_0 - \Ga_0 \Ga_{(j} \Ga_{i)}= 0$  lead
to the conclusion that the first term in  (\ref{cons1}) vanishes.
One can also easily see that the condition (\ref{cons2}) is
proportional to the LHS of (\ref{sym2}) and thus also vanish.
Finally, thanks to (\ref{sym1}) and (\ref{der}) one can see
that the first term of (\ref{cons3}) is equal to
$ a^{2}(2\theta J_{a} J^{a} + m \bar\psi \psi)$.

In summary this shows that  one can
express the energy-momentum tensor of the fermion in the
standard perfect-fluid form,
\beq
T_{\mu\nu} = (\rho + p) u_\mu u_\nu - p g_{\mu\nu}\,,
\eeq
where, for the FRW metric, \ $u_\mu = (a,\, 0,\, 0,\, 0)$.
Using identifications $T_{00} = \rho a^2$ and
$T_{ij}=p\,a^{2}  \eta_{ij}$, we are able to express
$\rho$ and $p$ as\footnote{Let us note that the zero
Energy-Momentum tensor without $\th$ in the massless
case is only due to the fact we perform this
calculation for a conformally-flat metric.
In this case the trace of the Energy-Momentum
tensor completely controls all its components. Some
extended discussion of this issue was recently given in
\cite{RadiAna}.}
\beq
\rho = \theta J^2 + m S
\qquad
{\rm and}
\qquad
p = \theta J^2\,.
\label{EOS}
\eeq
where $J^2 = J_\mu J^\mu = J^a J_a$.

The energy conservation law reads
\beq
\dot{\rho} \,+\,3H\, (\rho + p) \,=\, 0
\quad  \mbox{with} \quad H = {\dot a}/a\,,
\label{cons}
\eeq
This is consistent with the Dirac equations of motion derived previously.

For the FRW spacetime, every off-diagonal component of Einstein tensor
vanishes identically, \ $G_{0i} = G_{ij} = 0$ \ ($i\neq j$). This is
consistent with the form of the energy-momentum tensor just calculated
above. The temporal and spatial components of Einstein equations
(\ref{einstein}) read, respectively,
\beq
\frac{3a^{\prime 2}}{a^2}
& = & \frac{\ka}{2}a^2  \left( \theta J^2 + m S  \right)
=  \frac{\ka}{2}a^2 \rho \,, \label{einstein1} \\
-\frac{2a^{\prime\prime}}{a} + \frac{a^{\prime 2}}{a^2} & = &
\frac{\ka}{2}a^2  \theta J^2 =  \frac{\ka}{2}a^2 p \,, \quad
\mbox{.}
\label{einstein2}
\eeq
while taking the trace of the Einstein equation, we can express
the acceleration in terms of the physical time coordinate $t$  as
\beq
\frac{6\ddot{a}}{a} =  -\frac{\ka}{2} \left( 4\theta J^2+ mS \right)\,.
\label{ddota}
\eeq

\subsection{Massless case}

In the case where $m=0$ we can see that the BI parameter for a massless
Dirac field produces the same effect as of a perfect fluid characterized
by the equation of state (EOS) \ $p = \rho=\theta J^{2}$. It is remarkable
that, if $m = 0$, both energy density
and pressure in eq. (\ref{EOS}) do depend on $\th$, while their
ratio does not. Hence the theory under consideration does not
have a smooth limit for \ $\th\to 0$.
In this case the energy conservation law (\ref{cons})
gives us a new scaling law for energy density of massless
fermions in a gravity theory with non-zero BI parameter,
\beq
\rho \sim a^{-6}\,.
\label{rho}
\eeq

In the considerations presented above we did not take into
account the energy density and pressure of the free part of
the fermion spinor field. According to our calculations this
part is zero. However, this apparent result is only due to
our specific treatment of massless fermionic field in the
cosmological setting, where we are essentially looking for
the non-conformal part of $T_{\mu\nu}$. The conformal part
which does not couple to the conformal factor $a(t)$ can not
be seen in this approach for the following reason. We were
looking for the time-dependent spinor field, but this is
definitely not a right idea for the massless field which can
not be done space-independent by any choice of the local
reference frame. In order to see this, in the conformal
case, it is sufficient to consider the flat-space limit.
The space-independent fermion satisfies the equation
\beq
\Ga^0\pa_0\psi\,=\,0\,,
\label{N1}
\eeq
that simply means $\pa_0\psi=0$ and constant field. Then the Eqs.
(\ref{cons1}), (\ref{cons2}) and (\ref{cons3}) tell us that
$T_{\mu\nu}=0$. However the right way to take the energy density
and pressure in the conformal part of fermion $T_{\mu\nu}$ into
account is quite different. One has to remember that in the
massless field case the dependence on time and space coordinates
should be related, because the velocity is universally fixed for
all modes of the field. For example, the free solution of \ $i\Ga^\mu\pa_\mu\psi=0$ \ can be always considered as a
superposition of massless plane waves
\beq
\psi(x^0,x^i)\,=\,v\,e^{i(k_0x^0 - {\vec k} {\vec r})}
\,,\quad
k_0^2 - {\vec k}^2 = 0\,.
\label{N2}
\eeq
It is an elementary exercise to check that each one of these
waves creates pressure which is $1/3$ of its energy density.
Therefore the total pressure $p_0$ and energy density $\rho_0$
of the free fermionic field are related as $\,p_0\,=\,\rho_0/3$.
Taking into account our previous result we arrive at the
following energy density and pressure of the fermion field
in the presence of BI parameter:
\beq
\rho = \rho_0 + \theta J^{2}
\quad \mbox{and} \quad
p = p_0 + \theta J^2 \,=\,\frac13\,\rho_0 + \theta J^{2}\,.
\label{N4}
\eeq
Let us remark that the energy density and pressure of the
free radiation content of the Universe is not necessary
fermionic one, but can also include other fields, such as
electromagnetic radiation. The unique assumption which was
done here is that these two quantities are related by the
equation of state $\,p_0\,=\,\rho_0/3$, and this is indeed
satisfied for the asymptotically free ultra-relativistic
fields independent on their spin.

The Eqs. (\ref{N4}) indicate that the matter content of
the early Universe in the presence of the Holst term is
characterized by a sum of two fluids, one of them is
conventional radiation and another has equation of state
which is independent on the value of $\eta$. Of course,
the relevance of this second component depends on $\theta$,
but the equation of state does not.
In order to have better understanding of the new terms
with $\theta J^{2}$ in (\ref{N4}), we can consider the
simplified cosmological model with vanishing $\rho_0$ and
$p_0$. It is easy to solve the Friedmann equation, \
\ $H^2=(8\pi G/3)\,\rho$ \ and obtain the corresponding rule
for the expansion of the universe,
\beq
a \sim t^{1/3}\,,
\label{a}
\eeq
which, again, is different from the laws of expansion
for free radiation (with $a \sim t^{1/2}$) and dust (with
$a \sim t^{2/3}$). The results (\ref{rho}) and (\ref{a}) do
not depend on the value of BI parameter. The only one important
requirement is that this parameter should be nonzero. However,
the scaling law (\ref{rho}) shows that the new terms decay
much faster than radiation and hence become irrelevant in the
course of expansion of the Universe. At the same time, there
is a chance to see the traces of such term in observational
cosmological data, especially in the ones related to the cosmic
perturbations. In order to elaborate this idea one has to develop
the complete cosmological model with BI parameter. We postpone
this issue for the future work.

\subsection{Massive case}
In the massive case one has to distinguish the early and late time
regime. Let us recall that according to the Dirac equations we have
the conservation rule
\bea
J_{0}^{2} + P^{2} + S^{2} = \frac{M^{6}}{a^{6}},\qquad J_{i}^2
=\frac{\tilde{M}^{6}}{a^{6}}
\eea
where $M,\tilde{M}$ are  mass scales that control the size of the fermion
condensate.
This shows that $S$ is necessarily decreasing as the universe expand,
The late regime of the system is attained when
$$
a^{3}>> \frac{2\theta M^{3}}{m}.
$$
In this regime we have that
$2\theta S \ll m$ which means that we can neglect the non linear term
in Eq. (\ref{direq3}) and the equations for the evolution of $J_{0}$,
$S$, $P$ are dominated by the mass term; the influence of the $\th$
coupling can be ignored. We have, in this case,
\beq
\pa_{t}(a^{3} S) &\approx & 0\,,
\\
\pa_{t}(a^{3} J_{0}) &\approx & - 2m\,a^{3} P\,,
\\
\pa_{t}(a^{3} P) &\approx &  2m\, a^{3}J_{0}\,.
\eeq
This system of equation can be easily solved to give
\bea
J_{0} = \frac{M_{1}^3}{a^{3}} \cos\left[2m (t-t_{0}) \right]\,,
\quad
P = \frac{M_1^3}{a^{3}} \sin\left[2m (t-t_{0}) \right] \,,
\quad
S = \frac{M^3_2}{a^3}\,,
\eea
where $M_{1}$ and $M_{2}$ are mass scales characterizing the
condensate and satisfying $M^{6}= M_{1}^{6} + M_{2}^{6}$.
In this late time evolution the condensate evolve like a pressure
less fluid with $p=0$ where the energy density is dominated by
the scalar component and evolve like $\rho \sim a^{-3}$ while
$a \sim t^{2/3}$.

Another way to understand this result comes from the fact that  the
energy density  $\rho = \theta J^{2} + mS$ possess two components.
The scalar components scales like $ S\sim a^{-3}$ while the current
component scales like $J^{2}\sim a^{-6}$ therefore at late time the
energy contribution is dominated by the scalar component.

On the other hand this means that at an earlier time the current is
going to dominate the dynamic of the cosmological evolution. The
crossover time takes place when we can no longer neglect the
influence of the non linear $\theta$ term in the fermionic
evolution equation  (\ref{direq3}), that is when
\bea
a^{3} \sim \frac{2\theta M^{3}}{m}\,.
\eea
At this crossover time  both $mS$ and $\theta J^{2}$ are of the
same order of magnitude provided we assume that $M_{1}, M_{2}$
and $\tilde{M}$ are of comparable magnitude. The energy density
at this time is of the order $m^{2}/ \theta$ that is of the order
$ (\beta^{2}+1)/\beta^{2} m^{2}M_{P}^{2}$ where $M_{P}$ is the
Planck mass.

There are then two radically different early time evolution depending on
whether at this cross over time the current is space like or timelike.
Let us assume first that at this cross-over time the current is timelike:
i-e $J^{2}_{0}-J_{i}^{2} >0$ and that $\theta J^{2}$ is of the same order
as $ mS$.
In this case  the energy density is dominated at an earlier time by the
current term $\theta J^{2}$, the fermionic field becomes effectively
massless and the  dynamic for the fermionic equation of motion is
entirely dominated by the non linear term and given by the massless
equations:
\beq
\frac1{2a}(a^{3} S)^{\prime} &\approx & 2\theta J_{0}  (a^{3}P)\,,
\nonumber
\\
\frac1{2a}(a^{3} J_{0})^{\prime} &\approx & 0\,,
\nonumber
\\
\frac1{2a}(a^3 P)^{\prime} &\approx & - 2\th J_0 (a^3S)\,.
\label{eqs}
\eeq
In this regime applicable for early time
\bea
a^3 \ll \frac{2\theta M^{3}}{m}\,,
\eea
the solution reads
\bea
J_{0}=\frac{M^{3}_{1}}{a^{3}},\quad S =\frac{M^{3}_{2}}{a^{3}} \sin\left(
\int_{\eta_{0}'}^{\eta}\frac{2\theta M^{3}_{1}}{a^{3}} \right), \quad P
\,=\,
\frac{M^3_2}{a^3} \cos\left( \int_{\eta_{0}'}^{\eta}\,
\frac{2\theta M^{3}_{1}}{a^{3}} \right)\,.
\eea
where $M^{6}=M_{1}^{6} + M_{2}^{6}$.
In this early time regime the condensate evolves as massless condensate
$p=\rho$  with $\rho \sim a^{-6}$ while $a \sim t^{\frac13}$.
It is quite remarkable that the early time evolution although dominated
by the presence of $\theta$ is following an equation of state independent
of it.

Finally, there is another early time regime accessible to our system.
This regime happens if \ $J^{2}<0$, \ i.e., the current is space-like
at the cross-over time. In this case we cannot have that the current
term $\theta J^{2}$ dominates the scalar contribution since we should
respect the constraint that $\rho = \theta J^{2} + mS \geq 0$. What
happens in this case is that the quantity $\rho+p= 2\theta J^{2} + mS$
which is positive at late time decrease to eventually vanish.
When this quantity vanishes, and since $\dot{\rho}= -3H(\rho +p)$,
the energy density becomes constant. This means that we enter an
inflationary phase provided the value of the density energy is not
zero. This shows that the other regime of the theory, characterized
by a spacelike current at early time correspond to an inflationary
era that relax to pressure less dust at late time. The value of the
effective cosmological constant at early time depends on the initial
condition but it can be bounded by the value of $mS$ at crossover
over time. Indeed, since the inflationary era is characterized by
$2\theta J^{2}= - mS$, the effective cosmological constant is
$\rho_{\Lambda}= mS/2$. The value of $mS/2$ at early time is
necessarily smaller than its value at the cross over time which is
$m^{2}/2\theta^{2}$. So
\bea
\rho_{\Lambda} \leq \frac{m^{2}}{4 \theta^{2}}.
\eea

\section{Conclusions}

We have considered the cosmological solution for the
metric-spinor gravity with the Holst term, especially the
effect of the non-zero BI in a cosmological setting and shown
that  there are FRW-compatible solutions. One of the most
remarkable result is that the EOS of the self-interacting
spinor matter does not depend on the BI parameter in the massless 
limit, if we disregard the effect of free massless fermions. 
For the massive case we have identified two different regime.
In the first regime the theory is effectively massless at early
time and behaves as a perfect fluid with an ultra-hard equation
of state $w=1$. In the second regime the fermionic matter behaves
effectively as a cosmological constant and creates an inflationary
phase which is relaxed at late time into a pressureless fluid.

It would be definitely interesting to check whether the same
effects take place for the spinning fluid. In the positive case
this may lead to a potentially observable consequences for the 
early Universe. The effect of torsion and/or
Holst term on the EOS for the hot matter is depending on the
existence of axial current and on an arbitrary parameter $\th$
defined in Eqs. (\ref{th}) or (\ref{th-nonmi}). In principle,
some cosmological observations can be helpful in getting an upper
bound for $\th$. Let us note that recently the discussion of the
effect of Holst term and/or torsion on the difference between EOS
for photons and hot fermions (quarks and leptons) has been discussed
in \cite{Bojovald} in the framework of free fermion theory and loop
quantum gravity. The motivation for this study was to see whether
the fine balance required by Big Bang nucleosynthesis (BBN) holds
in the presence of loop quantum gravity effects. It was shown 
that the possible violation of such a fine balance due to the 
possible effects of torsion or Holst term is very small. 
In our understanding, the real physical situation can be even 
more simple, because fermions and photons are supposed to be 
interacting with each other at the BBN epoch. Such an interaction 
can control the balance between fermions and radiation through 
emission and absorbtion of photons by fermions and also by quantum 
effects such as creation and annihilation of fermion-antifermion 
pairs by radiation. As a result, the EOS for the matter content 
of the Universe should be treated as unique at that epoch and 
the effect of loop quantum gravity or torsion can not probably 
change, in principle, the fine balance between expansion 
rates of fermions and radiation. At the same time, this total 
EOS can be affected by the mentioned manifestations of a new 
physics, including the four-fermion interaction.  
\vskip 3mm

\noindent
{\bf Note added}. When we were preparing this manuscript for
submission, the preprint with partially similar content
\cite{Khriplovich} has been published.

\section*{Acknowledgements}

I.Sh. and G.B.P. are grateful to CNPq, FAPEMIG and CAPES
for partial support. C.A.S. is grateful to CAPES for the PhD
support program. L.F. acknowledges support from the Government
of Canada through Industry Canada and the Province of Ontario through
the Ministry of Research and Innovation.

\section*{Appendix A. \ Derivation of the Energy-Momentum Tensor}

For this end one has to take
\beq
g_{\mu\nu}\to g^\prime_{\mu\nu} = g_{\mu\nu} + h_{\mu\nu}
\,,\qquad
g^{\prime\mu\nu} = g^{\mu\nu} - h^{\mu\nu}
\,,\qquad
\sqrt{-g^\prime} = \sqrt{-g}\Big(1 + \frac{h}{2}\Big)\,,
\eeq
where only the contributions up to the first order in $h^{\mu\nu}$
were kept, and $h = h^\mu\mbox{}_\mu=h_{\mu\nu}g^{\mu\nu}$.
The expansion for the vierbeins are \cite{HD-psi}
\beq
e^{\prime b\al} = e^{b\al} - \frac{1}{2} h^\al\mbox{}_\be e^{b\be}
\,, \;\;\;\;\;\;
e^{\prime a}\mbox{}_\al = e^a\mbox{}_\al
+ \frac{1}{2} h_\al\mbox{}^\be e^a\mbox{}_\be\,,
\eeq
such that $J^{\prime\mu} J^\prime_\mu = J^\mu J_\mu$. This means that
$\de (J^\rho J_\rho)/\de g^{\mu\nu} = 0$.

In order to find the energy-momentum tensor coming from the kinetic
part of the Dirac action
\beq
S_k = \frac{i}{2}\int d^4x\sqrt{-g}{\cal L}_k
= \frac{i}{2}\int d^4x\sqrt{-g}\left\{
\bar{\psi}\gamma^\mu\nabla_\mu\psi -
\nabla_\mu\bar{\psi}\gamma^\mu\psi \right\}\,,
\eeq
we shall consider its variation in $h_{\mu\nu}$. By
straightforward calculations we obtain
\beq
S^\prime_k
& = & \frac{i}{2}\int d^4x \sqrt{-g}\Big(1 + \frac{h}{2}\Big)
\Big\{
\bar{\psi}\Big(\ga_\nu + \frac{1}{2}h^\rho\mbox{}_\nu\ga_\rho\Big)
\pa_\mu\psi (g^{\mu\nu} - h^{\mu\nu} )
\nonumber
\\
& - &
\pa_\mu\bar{\psi}
\left(\ga_\nu + \frac{1}{2}h^\rho\mbox{}_\nu\ga_\rho \right)
\psi (g^{\mu\nu} - h^{\mu\nu} )
+ \frac{i}{2}\bar{\psi}\Big(\ga_\nu
+ \frac{1}{2}h^\rho\mbox{}_\nu\ga_\rho \Big)
\om^{\prime ab}\mbox{}_\mu\si_{ab}\psi (g^{\mu\nu} - h^{\mu\nu})
\nonumber
\\
& + &
\frac{i}{2}\om^{\prime ab}\mbox{}_\mu\bar{\psi}\si_{ab}
\Big(\ga_\nu + \frac{1}{2}h^\rho\mbox{}_\nu\ga_\rho \Big)
\psi (g^{\mu\nu} - h^{\mu\nu} )\Big\}\,,
\eeq
where we used the covariant derivative of Dirac spinors in terms of the spin
connection, $\om^{ab}\mbox{}_\mu$,
\beq
\na_\mu\psi = \partial_\mu\psi
+ \frac{i}{2} \om^{ab}\mbox{}_\mu \si_{ab}\psi\,,
\quad
\na_\mu\bar{\psi} = \partial_\mu\bar{\psi}
- \frac{i}{2}\om^{ab}\mbox{}_\mu
\bar{\psi}\si_{ab}\,,
\eeq
and \ $\si_{ab} = \frac{i}{2}\,[\Ga_a\,,\Ga_b]$. Collecting the terms
up to first order in $h_{\mu\nu}$, and expressing the spin connection as
$$
\om^{\prime ab}\mbox{}_\mu = \om^{ab}\mbox{}_\mu
+ \Delta\om^{ab}\mbox{}_\mu\,,
$$
we find
\beq
S^\prime_k & = & S_k + \frac{i}{2}\int d^4x\sqrt{-g}\left\{
\frac{1}{2}g_{\mu\nu} {\cal L}_k -
\frac{1}{2}\left(\bar{\psi}\ga_{(\mu}\nabla_{\nu)}\psi
- \nabla_{(\nu}\bar{\psi}\ga_{\mu)}\psi\right)\right\} h^{\mu\nu}
\nonumber
\\
& + & \frac{i}{2}\int d^4x\sqrt{-g}\left\{
\bar{\psi}\ga^\mu\si_{ab}\psi + \bar{\psi}\si_{ab}\ga^\mu\psi\right\}
\Delta\om^{ab}\mbox{}_\mu\,. \label{Sk}
\eeq
It is easy to verify that $\Delta\om^{ab}\mbox{}_\mu$ can be written as
\beq
\Delta\om^{ab}\mbox{}_\mu = \frac{1}{4}\left\{\na_{\be}
\left( e^{a\al} e^{b\be} h_{\al\mu}\right) - \na_{\be}
\left( e^{b\al} e^{a\be} h_{\al\mu}\right)\right\}\,.
\eeq
Replacing this expression into equation (\ref{Sk}) and
performing integration by parts, one can conclude that
after some algebra all contributions
including $\Delta\om^{ab}\mbox{}_\mu$ do cancel identically
such that the momentum-energy tensor is given in Eq. (\ref{E}).

\section*{Appendix B. \  Equations of motion}

We start from the Dirac equation
\bea\label{Direq}
\frac{i}{a}\Gamma_{0} \left(
\psi^\prime +\frac{3a^\prime}{2a} \psi \right)
= m \psi + 2  \th \,(\Ga_5\Ga_{a} \psi) J^a\,.
\eea
The conjugate can be easily derived, but we do not present it here.
Let us remember the definition of the pseudo scalar
$P\equiv i \bar\psi \Ga_{5}\psi$, scalar  $S\equiv \bar\psi \psi$,
vector current $V_{a}= \bar{\psi}\Gamma_{a}\psi$, axial current
$J_{a}=\bar{\psi} \Gamma^{5}\Gamma_{a}\psi$, Lorentz  tensor
$L_{ab}= \frac{ i}{2} \bar\psi [\Ga_{a},\Ga_{b}]\psi$, the factors
of $i$ are chosen such that they are all real.

The first step is to establish the equation of motion for the scalar.
In order to do so we multiply (\ref{Direq}) by $\bar{\psi}\Gamma_{0}$
on the left and obtain
\beq
\frac{i}{a} \left(\bar{\psi}\psi^\prime
+ \frac{3a^\prime}{2a}\bar{\psi}\psi \right)& = &
 m \bar{\psi}\Gamma_{0}\psi
+ 2  \theta \, (  \bar{\psi} \Ga_{0} \Ga_5\Ga_{a} \psi ) J^a\,.
\label{N5}
\eeq

Since $\Ga_{5}= i \Ga_{0}\Ga_{1}\Ga_{2}\Ga_{3}$ we have that
$\Ga_{0}\Ga_{5}=  i \Ga_{1}\Ga_{2}\Ga_{3}$ and
$\Ga_{0}\Ga_{5} \Ga_{a}=  i \Ga_{1}\Ga_{2}\Ga_{3}\Ga_{a}
= - \eta_{a0} \Ga_{5} - i/2\epsilon_{aij} \Ga^{i}\Ga^{j}$.
Thus Eq. (\ref{N5}) becomes
\beq\label{scalar}
\frac{i}{a} \left(\bar{\psi}\psi^\prime
+ \frac{3a^\prime}{2a}\bar{\psi}\psi \right)& = &
 m V_{0}
+ 2\th\,\Big(i J_{0} P -\frac12 \epsilon_{aij}L^{ij} J^a\Big)\,,
\eeq
while its conjugate gives
\bea
-\frac{i}{a} \left(
\bar{\psi}^\prime\psi + \frac{3a^\prime}{2a}\bar{\psi}\psi \right)
= mV_{0}
+ 2 \theta\, \Big(-i J^0 P
-  \frac12\epsilon_{aij}L^{ij}J^a \Big)  \,.
\eea
By taking the imaginary part of (\ref{scalar}) we get
the conservation law
\bea
\frac{1}{2a^{4}}\left(a^{3} S \right)^{\prime} = 2  \theta\,  J_{0} P\,.
\eea
By taking the real part of (\ref{scalar}) we get
\bea
\frac{i}{2a} \left(\bar{\psi}\psi^\prime-\bar{\psi}^\prime\psi \right)
= mV_{0} -  \theta\, \epsilon_{aij}L^{ij} J^a\,.
\eea

Now we look at the pseudo scalar, by contracting  (\ref{Direq}) with
$\bar{\psi}\Gamma_{5}\Gamma_{0}$ we obtain:
\beq
\frac{i}{a} \left(\bar{\psi}\Ga_{5}\psi^\prime
+ \frac{3a^\prime}{2a}  \bar{\psi}\Ga_{5}\psi \right)
& = &  m \bar{\psi}\Ga_{5}\Gamma_{0}\psi
+ 2\th\,\big(\bar{\psi}\Ga_{5} \Gamma_{0}\Ga_5\Ga_{a} \psi \big)
\,J^a\,.
\eeq
Using $\Ga_{5}^{2}=1$ and $\bar{\psi}\Gamma_{0}\Gamma_{a}\psi =
\bar{\psi}\eta_{0a}\psi -i L_{0a}$, the RHS reads
$m J_{0}
- 2   \theta \, \big(J_{0}S  -i L_{0a} J^a\big)$.
Taking the real part we obtain
\bea
\frac{1}{2a^{4}}\left(a^{3} P \right)^{\prime} = m J_{0}
- 2   \theta \, J_{0}S.
\eea

We now consider the components of the axial vector by contracting
(\ref{Direq}) with $\bar{\psi}\Gamma_{5}$
\beq
\frac{i}{a}  \left( \bar{\psi}\Ga_5\Gamma_{0}\psi^\prime
+ \frac{3a^\prime}{2a} \bar{\psi} \Ga_5\Gamma_{0}\psi \right)& = &
 m   \bar{\psi} \Ga_5\psi
+ 2\th\,\big(\bar{\psi} \Ga_5 \Ga_5\Ga_{a} \psi \big) J^a\,.
\eeq
The RHS is $ -im P +2 \theta V_{a} J^{a}$, thus the imaginary part gives
\bea
\frac{1}{2a^{4}}\left(a^{3} J_{0} \right)^{\prime} = - m P.
\eea
We now look at the spatial components of the axial vector by contracting
(\ref{Direq}) with $\bar{\psi}\Gamma_{5}\Gamma_{i}\Gamma_{0}$:
\beq
\frac{i}{a} \left(\bar{\psi}\Ga_{5}\Ga_{i}\psi^\prime
+ \frac{3a^\prime}{2a}\bar{\psi}\Ga_{5}\Ga_{i}\psi \right)
& = &
 m \bar{\psi}\Ga_{5}\Ga_{i}\Gamma_{0}\psi
+ 2\th\,\big(\bar{\psi}\Ga_5\Ga_i\Ga_0\Ga_5\Ga_a\psi\big)J^a\,.
\eeq
Since $\Ga_{5} \Ga_{i}\Ga_{0}
=  - i/2\epsilon_{ijk} \Ga^{j}\Ga^{k}$, and
$\Gamma_{i}\Gamma_{0}\Gamma_{a}= \eta_{0a}\Gamma_{i}
-\eta_{ia}\Gamma_{0}
+ i\epsilon_{iab}\Gamma_{5} \Gamma^{b}$,
the RHS reads
$ -1/2 m \epsilon_{ijk} L^{jk} + 2  \theta \, (V_{i}  J_{0} -J_{i} V_{0}
+ i \epsilon_{iab} J^{a} J^{b}) $. Thus taking the imaginary part we get:
\bea
\frac{1}{a}\left(a^{3} J_{i} \right)^{\prime}
= 2\theta  \epsilon_{iab} J^{a} J^{b} =0\,.
\eea
We now consider the time component  vector current by contracting
(\ref{Direq}) with $\bar{\psi}$:
\beq
\frac{i}{a} \left( \bar{\psi}\Gamma_{0}\psi^\prime
+ \frac{3a^\prime}{2a} \bar{\psi} \Gamma_{0}\psi \right)
& = &
 m\bar{\psi}\psi
+ 2  \th\,\big(\bar{\psi} \Ga_5\Ga_a\psi\big) J^a\,.
\eeq
the RHS reads
$m S + 2\th\,J_a J^a$ which is real, thus we obtain
\bea
\frac{1}{a^{4}}\left(a^{3} V_{0} \right)^{\prime} =0,
\eea
while
\bea
\frac{i}{a} \left( \bar{\psi}\Gamma_{0}\psi^\prime
-  \bar{\psi}^{\prime}\Gamma_{0}\psi\right)
= m S + 2\th \,J_aJ^a\,.
\eea
The equation for the  space component   of vector current by contracting
(\ref{Direq}) with $\bar{\psi}\Gamma_{i}\Gamma_{0}$:
\beq
\frac{i}{a} \left(  \bar{\psi}\Ga_{i}\psi^\prime + \frac{3a^\prime}{2a}
\bar{\psi}\Ga_{i}\psi \right)
& = &  m\bar{\psi}\Ga_{i}\Gamma_{0}\psi
+ 2\th\, \big(\bar{\psi} \Ga_{i} \Ga_{0}\Ga_5\Ga_{a}\psi \big)J^a\,.
\eeq
Now  $ \Ga_{i} \Gamma_{0}\Ga_5\Ga_{a} = -\Ga_{i} \Gamma_{0}\Ga_{a} \Ga_5
=  -\eta_{0a} \Ga_{i}\Ga_5 + \eta_{ia} \Ga_{0}\Ga_5
- i\epsilon_{aij} \Ga^{j}$,
thus the RHS is
$-im L_{i0}  + 2\theta i\epsilon_{iaj} J^{a} V^{j}$.
Thus we have the evolution equation
\bea
\frac{1}{2a^{4}}\left(a^{3} V_{i} \right)^{\prime} = mL_{0i}
+ 2 \theta \epsilon_{iaj} J^{a} V^{j}
\eea
while
\bea
\frac{i}{a} \left( \bar{\psi}\Gamma_{i}\psi^\prime
-  \bar{\psi}^{\prime}\Gamma_{i}\psi\right)= 0.
\eea

The equation for the  space-time component   of Lorentz tensor is obtained
by contracting (\ref{Direq}) with $\bar{\psi}\Gamma_{i}$:
\beq
\frac{i}{a} \left(  \bar{\psi}\Ga_{i}\Ga_{0}\psi^\prime
+ \frac{3a^\prime}{2a} \bar{\psi}\Ga_{i}\Ga_{0}\psi \right)& = &
 m  \bar{\psi}\Ga_{i}\psi
+ 2\th\,\big(\bar{\psi} \Ga_{i} \Ga_5\Ga_{a}\psi\big) J^a  \,.
\eeq
Since $\Ga_{i}\Ga_{5}\Ga_{a}= -\eta_{ia}\Ga_{5}
- i \epsilon_{iab}\Ga^{0}\Ga^{b} +i/2 \eta_{a0}\epsilon_{ijk}
\Ga^{j}\Ga^{k}$ the RHS is
$mV_{i} + 2\theta(i J_{i} P - \epsilon_{iab}L^{0b}J^a
+ 1/2 \epsilon_{ijk} L^{jk}J_0)$.
Taking the real part of the equation we obtain
\bea
\frac{1}{2a^4} (a^{3} L_{i0})^{\prime} = m V_{i} - 2\theta \epsilon_{iab}
J^{a}L^{0b} +\theta J_{0}\epsilon_{ijk} L^{jk}.
\eea

The equation for the  space components   of the  Lorentz tensor
is obtained by contracting (\ref{Direq}) with
$\bar{\psi}\Gamma_{i}\Ga_{j}\Ga_{0}$:
\beq
\frac{i}{a} \left(  \bar{\psi}\Ga_{i}\Ga_{j}\psi^\prime
+ \frac{3a^\prime}{2a} \bar{\psi}\Ga_{i}\Ga_{j}\psi \right)& = &
 m  \bar{\psi}\Gamma_{i}\Ga_{j}\Ga_{0}\psi
+ 2  \th\, \big(\bar{\psi} \Ga_{i}\Ga_{j}\Ga_{0}
\Ga_5\Ga_{a} \psi \big) J^a  \,.
\eeq
Using that $\Gamma_{i}\Ga_{j}\Ga_{0} = \eta_{ij}\Gamma_0
-i\epsilon_{ijk}\Ga_{5}\Ga^{k}$ and that
\beq
\Ga_{i}\Ga_{j}\Ga_{0} \Ga_5\Ga_{a} =
\frac{i}{2}\big(\eta_{ai}\epsilon_{jbc} -
\eta_{aj}\epsilon_{ibc}\big)\Ga^{b}\Ga^{c}
- i\eta_{a0} \epsilon_{ijk}\Ga^{0}\Ga^{k} + i\epsilon_{ija}
- \eta_{0a}\eta_{ij}\Gamma_5
- \frac{i}{2} \eta_{ij}\epsilon_{akl}\Gamma^k\Gamma^l\,,
\nonumber
\eeq
we obtain, antisymmetrizing in $i$ and $j$,
\bea
\frac{1}{2a^{4}} (a^{3} L_{ij})^{\prime}
= 2\theta\Big( \frac12 J_{i}
\epsilon_{jab}L^{ab} - \frac12 J_{j}\epsilon_{iab}L^{ab} - J_{0} \epsilon_{ijk}
L^{0k}\Big).
\eea
Contracting this with $\epsilon^{ijk}/2 $ gives
\bea
\frac{1}{2a^{4}} (a^{3} L^{k})^{\prime}  =2\theta \left( \epsilon^{ijk}
J_{i}L_{j} + J_{0}L^{0k}\right)\,.
\eea



\end{document}